\documentclass[11pt]{article}

\usepackage{amsfonts,amssymb,amsmath,amsthm,geometry}
\usepackage{graphicx,color}
\usepackage{subfigure}
\usepackage{float}

\def\beq#1#2\eeq{\begin{equation}\label{#1}#2\end{equation}}
\def\bal#1#2\eal{\begin{align}\label{#1}#2\end{align}}
\def\bse#1#2\ese{\begin{subequations}\label{#1}#2\end{subequations}}
\def\ba{\begin{aligned}}
\def\ea{\end{aligned}}

\def\dd{\operatorname{d}}

\begin{document}

\Large 
\begin{center}
{Comment on "Characterization of an acoustic spherical cloak", \\ {\it Inverse Problems}, {\bf 31}(3):035001, 2015.\footnote{This Comment was     rejected by {\it Inverse Problems}.}
}
\end{center}
\normalsize \rm

\vspace{3mm}

\large
\centerline{A. N. Norris}
\normalsize 

\vspace{3mm}

\centerline{Mechanical and Aerospace Engineering, Rutgers University, Piscataway, USA  }

\normalsize

%
%

\bigskip
The paper \cite{Dassios2015} considers a  
spherical cloak described by three radially varying acoustical  quantities. 
For a given radial mass density in the cloak the question  posed is whether the remaining two parameters, tangential density and compressibility, are uniquely determined.   A  
 method is proposed in  \cite{Dassios2015} to solve this inverse question  based upon  the solution of  a Riemann-Hilbert problem involving spectral properties of a one-dimensional inhomogeneous Schr\"odinger equation.  However,  no constructive examples of the solution procedure are given. 
This comment provides  explicit  solutions for any radial mass density that conforms with the requirements of transformation acoustics \cite{Norris08b}.  A valid form  of the compatibility condition \cite[eq.\ (22)]{Dassios2015} for $n=0$ is derived.

 \section{Explicit solutions}

The problem concerns a spherical shell $V$$: a<r<b$, the exterior infinite  region $V^e$$: b<r$, and the interior $V^i$$: 0<r<a$.  Both $V^e$ and $V^i$ are occupied by   acoustic fluids of uniform mass density, compressibility  and sound speed equal to 
$\{ \rho_0, \, \gamma_0, c= (\rho_0 \gamma_0)^{-1/2}\}$ and 
$\{ \rho_1, \, \gamma_1, c_1= (\rho_1 \gamma_1)^{-1/2}\}$, respectively ({\cite{Dassios2015} considers the case $\rho_1=\rho_0$, $\gamma_1=\gamma_0$.).  
  The cloaking domain $V$ has radially varying parameters, 
radial density $\rho (r)$,  tangential density $s(r)$ and compressibility $\gamma (r)$.   The objective is to ensure zero scattering for plane wave incidence.  Time harmonic dependence $e^{-i\omega t}$ is assumed.  Let $u$ denote the acoustic pressure field, then in $V^e$  
$u=  u^e ({\bf r}) \equiv e^{i {\bf k}\cdot{\bf r}}$ where ${\bf k}= k\hat {\bf k}$ is the wavenumber vector of magnitude $k=\omega /c$.

\subsection{ $s(r)$ and $\gamma(r)$ in terms of $\rho(r)$}

Consider eq.\ (19) of \cite{Dassios2015}, replacing $f_n^m$ by $f_n$ since the solutions are independent of $m$, 
\beq{1}
\frac{\dd}{\dd r}\Big(
\frac{r^2} {\rho(r)}
\frac{\dd f_n(r)}{\dd r}
\Big)
+ \Big(
r^2 \gamma(r) \omega^2 - 
\frac{n(n+1)}{s(r)} 
\Big) f_n(r) =0, \ \  a<r<b .
\eeq
For a given positive radial density $\rho(r)$, define 
\beq{3}
s(r)=   \frac{\rho_0^2 r^2}{\rho(r)R^2(r)}   , 
\ \ 
\gamma(r)  = \frac {\rho_0^3 \gamma_0} { \rho(r) s^2(r)}    
\ \ 
\eeq
where
\beq{2+}
R(r) =  \Big( \frac 1b + \int_r^b  \frac{\rho(t)}{\rho_0 t^2} \dd t
\Big)^{-1}. 
\eeq 
The dependence $r\to R(r)$ is one-to-one since $\rho(r) $ is positive and hence $R$ can be considered a mapping or transformation.  
Equation \eqref{1}  becomes 
\beq{5}
\frac{\dd}{\dd R}\Big(
{R^2} 
\frac{\dd F_n(R)}{\dd R}
\Big)
+ \big( k^2 R^2   - {n(n+1)}
\big) F_n(R) =0, \ \  R(a) <R<b 
\eeq
where 
\beq{2}
F_n(R) = f_n (r) .
\eeq 
The unique solution of \eqref{1} satisfying the boundary conditions eqs.\ (20) and (21) of \cite{Dassios2015}
\beq{5.1}
f_n(b) = j_n(k b),
\ \ 
\frac 1{\rho(b)} f_n ' (b) = \frac k{\rho_0} j_n '(k b)
\eeq
is therefore 
\beq{6}
f_n(r) = j_n(k R(r)) .
\eeq

{\bf Example 1.}  The radial density of  \cite{Cummer08} is 
\beq{3.1} 
\rho(r) = \frac{b-a}b \Big(\frac r{r-a}\Big)^2 \rho_0 , 
\eeq
for which eqs.\  \eqref{3} and \eqref{2+} give  
\beq{3.2} 
R(r) = \frac{r-a}{b-a}b , 
\ \ 
s(r) = \frac{b-a}b   \rho_0 , 
\ \ 
\gamma(r) = \Big( \frac{b}{b-a} \Big)^3 \Big( \frac{r-a}{r} \Big)^2 \gamma_0 .
\eeq
Note that $\gamma(r)$ follows from  \eqref{3}$_2$, i.e. the identity 
$\rho (r) s^2(r) \gamma(r) = \rho_0^3 \gamma_0$.
In further examples $\gamma(r) $ is not given explicitly. 

{\bf Example 2.}   The linear near-cloak of  \cite{Kohn08} corresponds to 
\beq{3.3} 
\rho(r) = \frac{b-a}{b-\delta} \bigg(
\frac r{  r - \big(\frac{a-\delta}{b-\delta}\big) b}\bigg)^2 \, 
 \rho_0 , 
\eeq
for which eqs.\  \eqref{3} and \eqref{2+} yield   
\beq{3.4} 
R(r) = 
\frac{r-a}{b-a}b + \frac{b-r}{b-a}\delta  
, 
\ \
s(r) = \frac{b-a}{b-\delta}   \rho_0 .
\eeq
This example reduces to Example 1 if $\delta = 0$. 

{\bf Example 3.} 
Constant radial density $\rho(r) = \rho_c$ yields 
\beq{4.0}
R(r) = \Big( \frac 1b +    \frac{\rho_c}{\rho_0 } \big( \frac 1r - \frac 1b\big)
\Big)^{-1}, 
\ \ 
s(r) = \frac{r^2 \rho_0^2}{R^2(r)\rho_c} . 
\eeq

{\bf Example 4.}  
In general, for a given function $R(r)$  \cite[eq.\ (2.16)]{Norris08b}, 
\beq{4.1}
\rho (r) = \frac{\rho_0 \, r^2}{R^2(r)} R'(r)  , 
\ \ 
s(r) =  \frac { \rho_0}{R'(r)} , 
\ \ 
\gamma(r) =  \frac{R^2(r)}{r^2}  {R'(r)}  \gamma_0 . 
\eeq

The parameters $\{\rho(r),\, s(r),\, \gamma(r)\} $ correspond respectively to 
$\{ 1/K_r(r),\, 1/K_\perp(r),\, \rho(r)\} $ of \cite{Gokhale2012} where other examples are considered.  

\section{Compatibility condition}

The solution $f_n(r)$, $n\ge 0$,  must satisfy a  compatibility condition at $r=a$ if the interior region is to be cloaked.  We revisit this  because of the more general problem considered here of an interior region with different properties  than the exterior,  and also because the compatibility condition \cite[eq.\ (22)]{Dassios2015} is not valid for $n=0$.   A corrected compatibility condition for $n=0$ will be  derived. 

The pressure in $V^i$ is 
$u^i({\bf r})$ which 
satisfies the Helmholtz equation \cite[eq.\ (2)]{Dassios2015}
\beq{2.4}
\Delta u^i({\bf r}) + k_1^2 u^i({\bf r}) =0,  \ \ {\bf r}\in V^i  
\eeq 
where $k_1=\omega /c_1$. 
The particle velocity follows from \cite[eq.\ (7)]{Dassios2015}  as 
\beq{2.2}
{\bf v}=   (i\omega \rho_1 )^{-1} \nabla u^i 
, \ \ {\bf r}\in V^i.
\eeq
The solutions in the core and the shell are \cite[eqs.\ (17), (18)]{Dassios2015}
\beq{2.5}
\left. 
\begin{matrix} u^i  ({\bf r})  \\ u  ({\bf r}) 
\end{matrix}
\right\}
= 4\pi \sum_{n=0}^\infty \sum_{m=-n}^{m=n} i^n Y_n^m( \hat{\bf r}) Y_n^m( \hat{\bf k})^*
\, \times\,
\begin{cases}
A_n^m j_n(k_1r) , & {\bf r}\in V^i, 
\\
f_n^m(r)  , & {\bf r}\in V . 
\end{cases}
\eeq
The continuity conditions for pressure and radial velocity at the interface $r=a$ \cite[eqs.\ (8), (9)]{Dassios2015} therefore become
\beq{2.6}
\begin{aligned}
f_n^m(a) &= A_n^m j_n(k_1a)  , 
\\
\frac 1{\rho(a)} {f_n^m}' (a) &= \frac {k_1}{\rho_1} A_n^m j_n' (k_1a)  
. 
\end{aligned}
\eeq
Eliminating $A_n^m $ and dropping the redundant superscript $m$ yields the compatibility condition 
\beq{2.7}
   k_1\rho(a) j_n' (k_1a) f_n(a) =  {\rho_1} j_n(k_1a) {f_n}' (a)  
\eeq
which is the generalized form of \cite[eq.\ (22)]{Dassios2015}.   

The compatibility condition  is satisfied by the solution \eqref{6} iff  
\beq{9}
j_n'(k_1 a) j_n(k R(a)) - 
\frac{R^2(a)}{a^2}  \frac{\rho_1 c_1}{\rho_0 c} j_n(k_1 a)  j_n'(k R(a)) 
   = 0. 
\eeq
This holds $\forall \, n \ge 1$ if 
\beq{10}
R(a)=0
\eeq
which is the condition expected from transformation acoustics \cite{Norris08b}. 
Using the identity \eqref{10} the $n=0$ compatibility condition \eqref{9} reduces to 
\beq{101}
-j_1(k_1 a)  = 0, 
\eeq
which only holds  if $k_1 a$ is a zero of $j_1$.    Hence, the 
$n=0$  compatibility condition does not appear to be correct since it is not satisfied even for vanishing $R(a)$, which is sufficient according to transformation acoustics.   The same inconsistency  applies to the compatibility condition \cite[eq.\ (22)]{Dassios2015} for the radial  density  considered in Example 1 of \cite{Dassios2015}, i.e. $\rho(r)$ of   \cite[eq.\ (27)]{Dassios2015} which is the same as $\rho(r)$ of eq.\ \eqref{3.1} above.  Thus, using this $\rho(r)$, along with 
$f_n(r) = j_n\big( kb (r-a)/(b-a)\big)$  
\cite[eq.\ (32)]{Dassios2015} in eq.\ \eqref{2.7} (or equivalently \cite[eq.\ (22)]{Dassios2015}) for $n=0$ yields zero on the right hand side and infinity on the left.  Furthermore, this indicates that the value of $A_0$ in \cite[eq.\ (34)]{Dassios2015} is suspect. 
The proper form of the $n=0$ compatibility condition and the correct value of $A_0$ are examined next. 

\subsection{A compatibility condition for all $n\ge 0$}

In order to derive a consistent   compatibility condition that holds for all values of $n \ge 0$ we    
introduce waves going  both ways in $V$ and $V^e$.  
The solution in $V^i$ is still given by  \eqref{2.5}, while the  solutions  in the cloak and exterior domains become
\beq{2.51}
\left. 
\begin{matrix} u^e  ({\bf r})  \\ u  ({\bf r}) 
\end{matrix}
\right\}
= 4\pi \sum_{n=0}^\infty \sum_{m=-n}^{m=n} i^n Y_n^m( \hat{\bf r}) Y_n^m( \hat{\bf k})^*
\, \times\,
\begin{cases}
  \, \big( j_n(kr)  + B_n^m h_n(kr)  \big)  , & {\bf r}\in V^e, 
\\
\, \big( j_n(k R(r))  + B_n^m h_n(k R(r))  \big)  , & {\bf r}\in V . 
\end{cases}
\eeq
where $h_n = h_n^{(1)}$ is the spherical Hankel function of the first kind.   Continuity of pressure and radial velocity at $r=b$ is automatically satisfied. 
The interface conditions at $r=a$ become (dropping the superscript $m$)
\beq{5.7}
\begin{aligned}
  j_n(k R(a))+ B_n h_n(k R(a)) &= A_n j_n(k_1a), 
\\
\frac{\rho_1c_1}{\rho_0 c} \epsilon^2 \big( j_n'(k R(a))+ B_n h_n'(k R(a))\big) &= A_n j_n'(k_1a) , 
\end{aligned}
\eeq
with solutions,  
\bal{5.8}
A_n &=  \frac
{ - i\frac{\rho_1c_1}{\rho_0 c} (k a)^{-2} }
{h_n(k R(a)) j_n'(k_1a) - \frac{\rho_1c_1}{\rho_0 c} \frac{R^2(a)}{a^2} h_n'(k R(a)) j_n(k_1a) }
, 
\\
B_n &= -\bigg(\frac
{j_n(k R(a)) j_n'(k_1a) - \frac{\rho_1c_1}{\rho_0 c} \frac{R^2(a)}{a^2} j_n'(k R(a)) j_n(k_1a) }
{h_n(k R(a)) j_n'(k_1a) - \frac{\rho_1c_1}{\rho_0 c} \frac{R^2(a)}{a^2} h_n'(k R(a)) j_n(k_1a) }
\bigg) 
\eal 
where $j_n(x)h_n'(x) - j_n'(x)h_n(x) = i x^{-2}$ has been used. 
As a compatibility condition we require
\beq{3=6}
B_n =0  
\eeq
  which is satisfied by $R(a)=0$, i.e.\ eq.\ \eqref{10}, for all $n\ge 0$.  Furthermore, taking the limit of 
	\eqref{5.8} as $R(a) \to 0$ yields 
\beq{88}
 A_n  = 0 ,  \ \ n\ge 0 .
\eeq
In summary, the compatibility condition \eqref{3=6}, which is valid for all $n\ge 0$, replaces the  condition \eqref{9} (and \cite[eq.\ (22)]{Dassios2015}), and eq.\ \eqref{88} corrects eq.\ (34) of  \cite{Dassios2015}.

\section{Conclusion}

There is an infinite set of radial density functions $\rho (r)$ for which the tangential density $s(r)$ and  compressibility $\gamma(r)$ are given by eqs.\ \eqref{3} and \eqref{2+}.  These solutions require that $R(a) = 0$ which in turn means that $\rho (r)$ must be singular as $r\downarrow a$.   Define the radial "mass" of the cloak  enclosed between  $r$ and $b$, 
\beq{56}
m(r) = 4\pi \int_r^b \rho(r)r^2 \dd r   ,
\eeq
which  can be expressed as 
\beq{57}
m(r) = 4\pi  \rho_0 a^4 \Big( \frac 1{R(r) } - \frac 1b\Big) + 
4\pi \int_r^b \big( \frac{r^4-a^4}{r^2}\big) \rho(r)\dd r . 
\eeq
The latter integral is positive while the term  $ 1/{R(r) }\to  \infty $ as $r\to a$. Hence,  the total radial mass of the cloak must be infinite.  This conundrum was  noted in \cite{Norris08b} where an alternative solution involving anisotropic stiffness and isotropic density was proposed, and was shown to have finite mass equal to the mass of fluid inside a sphere of radius $b$. 

The question regarding \cite{Dassios2015} is whether other solutions exist  for the positive triplet
$\rho(r),\, s(r),\, \gamma(r)$, specifically solutions for which $\rho(a)$ is finite.   Intuition suggests that the answer is no, that the only solutions correspond to the transformation function $R(r)$ mapping the inner boundary $r=a$ to the origin.  A counterexample  would not only be very interesting, but would  provide new directions for research: one can only hope that the  algorithm developed in  \cite{Dassios2015} can provide some insight.

\end{document}